\documentclass{article}

\usepackage{arxiv}

\usepackage[utf8]{inputenc} % allow utf-8 input
\usepackage[T1]{fontenc}    % use 8-bit T1 fonts
\usepackage{hyperref}       % hyperlinks
\usepackage{url}            % simple URL typesetting
\usepackage{booktabs}       % professional-quality tables
\usepackage{amsfonts}       % blackboard math symbols
\usepackage{nicefrac}       % compact symbols for 1/2, etc.
\usepackage{microtype}      % microtypography
\usepackage{lipsum}
\usepackage{graphicx}

\title{Assessing the Intrinsic Uncertainty and Structural Stability of Planetary Models: 1) Parameterized Thermal-Tectonic History Models}

\author{
  Johnny Seales \\
  Rice University \\
  \texttt{jds16@rice.edu}
   \And
  Adrian Lenardic \\
  Rice University \\
  \texttt{ajns@rice.edu}\\
  \AND
  William moore \\
  Hampton University \\
  \texttt{william.moore@hamptonu.edu} \\
}

\begin{document}
\maketitle

\begin{abstract}
Thermal history models, that have been used to understand the geological history of Earth, are now being coupled to climate models to map conditions that allow planets to maintain surface water over geologic time - a criteria considered crucial for life. However, the lack of intrinsic uncertainty assessment has blurred guidelines for how thermal history models can be used toward this end. A model, as a representation of something real, is not expected to be complete. Unmodeled effects are assumed to be small enough that the model maintains utility for the issue(s) it was designed to address. The degree to which this holds depends on how unmodeled factors affect the certainty of model predictions. We quantify this intrinsic uncertainty for several parameterized thermal history models (a widely used subclass of planetary models). Single perturbation analysis is used to determine the reactance time of different models. This provides a metric for how long it takes low amplitude, unmodeled effects to decay or grow. Reactance time is shown to scale inversely with the strength of the dominant feedback (negative or positive) within a model. A perturbed physics analysis is then used to determine uncertainty shadows for model outputs. This provides probability distributions for model predictions and tests the structural stability of a model. That is, do model predictions remain qualitatively similar, and within assumed model limits, in the face of intrinsic uncertainty. Once intrinsic uncertainty is accounted for, model outputs/predictions and comparisons to observational data should be treated in a probabilistic way.
\end{abstract}

\section{Introduction}
\indent The discovery of terrestrial exoplanets has rejuvenated interest in determining the factors that maximize the potential for life on a planet \citep[e.g.][]{Meadows2018}. Modeling the evolution of terrestrial planets, with a focus on mapping the range of conditions that can maintain life potential, falls under this research umbrella. Consideration of solar energy feeds into this modeling exercise, as does a consideration of a planet's internal energy. Internal energy drives volcanic and tectonic activity. This, in turn, cycles planetary volatiles between surface and interior envelopes and volatile cycling may be critical for maintaining clement conditions over timescales that allow life to develop and evolve \citep{Walker1981,Kasting1993}. The internal energy of terrestrial planets decays over time and the interior of a planet, in the long term, cools until it becomes geologically inactive. Determining/mapping particular and/or potential planetary cooling paths is the goal of  thermal history modeling \citep{Davies1980,Schubert1980}. Thermal history models developed for the Earth have been, and continue to be, adopted and adapted for exoplanet modeling \citep{Kite2009,Schaefer2015}. They are also being linked to climate models so as to model the surface conditions of terrestrial planets over geological time scales \citep{Jellinek2015,Foley2015,Lenardic2016,Foley2018,Rushby2018}.

Thermal history modeling seeks to map the cooling path of a particular body or, for comparative studies, to map cooling curves for different planets and/or moons. Sub-solidus thermal convection in the rocky interior (the mantle) of terrestrial planets is a key cooling mechanism in their evolutions \citep{Schubert1979}. Modeling mantle convection over the geologically active lifetime of a terrestrial planet is, in principal, possible using 3D numerical models that solve the coupled equations of heat and mass transport. In practice, this approach has restrictive limits. For an Earth sized planet, the vigor of thermal convection in its early evolution pushes the resolution limits needed for numerical accuracy beyond the edge of current computational power. Such models can be run over truncated thermal history times but they remain time consuming. Using them to map a wide range of potential cooling paths, associated with uncertainty in material properties and/or initial conditions, remains out of reach under practical time constraints. All of these issues become more extreme for modeling the evolution of large terrestrial exoplanets of the type that have been discovered in solar systems beyond our own.

For these reasons, simplified thermal evolution models -- that track spatially averaged quantities -- remain popular for both Earth-focused studies \citep{Conrad1999, Korenaga2003, Sandu2011, Foley2012, Hoink2013} and for exoplanet studies \citep{Kite2009, Schaefer2015, Komacek2016}. The reduced models are referred to as parameterized thermal history models since they do not solve the full convection equations but use, instead, an empirical formulation that parameterizes variations in convective heat flux as a function of the physical factors that drive and resist convective motions \citep{Schubert1979,Schubert1980}. These types of models are efficient for traversing vast regions of parameter space \citep{McNamara2000}. They also allow layers of complexity to be added to base level models in relatively simple and efficient ways - for example: deep water cycling \citep{McGovern1989,Sandu2011}, planetary carbon cycling \citep{Tajika1990,Tajika1992,Franck1995,Sleep2001,Sleep2001a,Abbot2012}, coupled thermal evolution and climate modeling \citep{Jellinek2015,Foley2015,Lenardic2016}. A final advantage for exoplanet modeling is that they can be scaled to different planetary mass and/or volume in ways that maintain model efficiency \citep{Valencia2006}.

Informed application and/or evaluation of planetary models (thermal history models being an example) requires assessment of model uncertainties. A defining feature of a model is that it is, at some level, simpler than the phenomena it seeks to model. This introduces an {\em intrinsic} uncertainty associated with the effects of unmodeled factors (both known and unknown). This type of uncertainty is also referred to epistemic or systemic uncertainty. We use the term {\em intrinsic} herein to highlight that this type of uncertainty is connected to the very definition of a model - it is intrinsic to the modeling process. There is also uncertainty associated with the fact that planetary models involve multiple input parameters that are not perfectly known (e.g., material properties, initial conditions). Parameter uncertainty can be assessed through sensitivity analysis: a number of models are run with variable input conditions to determine how model outputs respond \citep[e.g.][chap. 9]{Saltelli2008,Loucks2017}. Intrinsic uncertainty is less often directly assessed but a general assumption is that the effects of model simplifications do not alter the ability to make useful comparisons of model outputs to observations (i.e., the intrinsic uncertainty is assumed to be negligible compared with the uncertainty associated with observational data and/or with input parameter uncertainties).  \\

Modeling the Earths thermal history has focused on models that can match observations. Constraints from Earth observations are often built directly into thermal history models \citep{Christensen1984,Korenaga2003}. The intrinsic uncertainty of the models, when used in Earth applications, is rarely quantified. To the best of our knowledge, there is no published study that has directly evaluated the intrinsic uncertainty of a thermal history model. When observational data is available, the proof of a model's worth is often seen exclusively as its ability to account for the data.  If a model can not match the data then it is considered incorrect in terms of assumptions and/or incomplete in the sense that unmodeled factors are not negligible. The latter can be seen as an indirect assessment of intrinsic uncertainty. However, that type of approach can mix intrinsic and parameter uncertainty in a manner that masks the quantitative aspects of the former. Uncertainty in the observational data itself can add to the screening of intrinsic model uncertainty if the only assessment used is how well model results match data. 

We can phrase it another way: models applied to Earth data are postdictive. That is, they seek to provide explanations for existing data. In the process, uncertainty can be hidden as the models are effectively fit to Earth data, i.e., the tuning of model parameters to match Earth observations can screen intrinsic model uncertainties.  Even if a thermal history model can match Earth observations, the model will still have an intrinsic uncertainty. The lack of quantitative uncertainty measures can become problematic when the models are shifted into a predictive mode for exoplanet application. The goal of this work is to quantify the intrinsic uncertainty of different thermal history models so that the uncertainties between different models, and between model outputs and data can be quantitatively compared. \\

Intrinsic uncertainty is also referred to as structural error \citep{Strong2014,Wieder2015a}. Model structure refers to the specific cause and effect relationships that define a model. The outputs/solutions from a structurally stable model do not change qualitatively if the model is perturbed slightly. If, on the other hand, small amplitude perturbations cause the qualitative nature of model solutions to change (e.g., attractors in model solution space appear or disappear), then the model is structurally unstable \citep{Guckenheimer1983,George1985}. Testing the structural stability of a model can also provide an assessment of intrinsic uncertainty. Even if a model is structurally stable to low amplitude perturbations, its outputs will change quantitatively and that change can provide an uncertainty measure. The remainder of this paper develops and assesses intrinsic uncertainty measures for a range of thermal history models that have been applied to the Earth. All of the models assessed assume that the geological manifestation of mantle convection is in the form of plate tectonics. However, the methods we employ can be applied to models that assume other forms of planetary tectonics \citep{Lenardic2018}. \\

\section{METHODS}
\subsection{Paramterized Thermal History Models}
\indent Parameterized thermal history models are based on a balance between the heat generated within the interior of a terrestrial planet's mantle and the heat flow through the planet's surface according to
\begin{equation} \label{Tdot}
C\dot{T}=H-Q
\end{equation}
where $C$ is the heat capacity of the mantle, $\dot{T}$ is the change in average mantle temperature with time, $H$ is the total amount of heat produced internally and $Q$ is the total surface heat flow. \\

For most of the Earth's history, heat is produced principally by the radiogenic decay $^{238}$U, $^{235}$U, $^{232}$Th and $^{40}$K isotopes. The total amount of heat produced at any time is modeled as
\begin{equation} \label{H}
H(t)=H_0\sum_{n=0}^{4} h_nexp\left(\lambda_nt\right), 
h_n=\frac{c_np_n}{\sum_nc_np_n}
\end{equation}
where $H_0$ is total present day heat generation, $h_n$ is the amount of heat produced by a given isotope, $\lambda_n$ is the decay constant for a given isotope, and $t$ is time. Present day fractional concentrations and power production for each isotope are represented by $c_n$ and $p_n$, respectively. The values used for each parameter is given in Table \ref{table:radiogenics}. We assume present day proportions of $U:Th:K=1:4:(1.27x10^4)$ and normalize by total U to calculate relative isotope concentrations. Although we will not include them here, short-lived isotopes of Al and Fe can also make significant contributions to internal heat early in the history of rocky planets \citep{MacPherson1995}. \\ 

\begin{table}[h]
\caption{Radiogenic Heat Production}
\centering
\begin{tabular}{ c c c c c }
\hline\hline
Isotope & $P_n$ $(W/kg)$ & $c_n$ & $h_n$ & $\lambda_n$ $(1/Ga)$ \\
\hline
$^{238}U$  & $9.37*10^{-5}$ & 0.9927 & 0.372  & 0.155 \\
$^{235}U$  & $5.69*10^{-4}$ & 0.0072 & 0.0164 & 0.985 \\
$^{232}Th$ & $2.69*10^{-5}$ & 4.0    & 0.430  & 0.0495 \\
$^{40}K$   & $2.79*10^{-5}$ & 1.6256 & 0.181  & 0.555 \\
\hline
\end{tabular}
\label{table:radiogenics}
\end{table}

The heat flux through the surface, which depends on convective vigor in the mantle, is typically parameterized using a scaling equation given by \citep{Schubert2001}: 
\begin{equation}\label{eqparam}
Nu=aRa^{\beta}
\end{equation}
where $a$ is a constant, $Nu$ is the Nusselt number defined as the  convective heat flux normalized by the amount of heat that would be conducted over the entire layer. The Rayleigh number ($Ra$) is  defined as
\begin{equation} \label{Ra}
Ra=\frac{\rho g\alpha\Delta T D^3}{\kappa \eta (T_m)}
\end{equation}
where $\rho$, $g$, $\alpha$, $\Delta T$, $D$,$\kappa$, and $\eta (T_m)$ are density, gravity, thermal expansivity, the depth of the convecting mantle, the  difference between surface and internal mantle temperature, thermal diffusivity and viscosity, respectively. The scaling parameter, $\beta$, varies for different classes of thermal history models. We will return to this issue shortly. \\

The temperature-dependent viscosity of the mantle is defined as
\begin{equation} \label{Visc}
\eta(T_m)=\eta_0exp\left(\frac{A}{RT_m}\right)
\end{equation}
where $A$ is the activation energy, $R$ the universal gas constant, and $\eta_0$ a constant \citep{Karato1993}. As temperature increases, the viscosity will decrease, leading to an increase in $Ra$. If we assume that all values in equation (\ref{Ra}) are constant except $T_m$ and $\eta (T_m)$, then combining equations (\ref{eqparam}), (\ref{Ra}), and the definition of $Nu$ leads to
\begin{equation} \label{Q}
Q=a'\frac{T_m^{1+\beta}}{\eta (T_m)^{\beta}}
\end{equation}
where all constants have now been combined into $a'$. \\

Classic thermal history models, developed for Earth, set $\beta$ to a value of 0.33 based on laboratory experiments and boundary layer theory \citep[e.g.][]{Davies1980,Schubert1980}. The constant, $a$, in equation (\ref{eqparam}), is determined based on experimental results or numerical simulations. The viscosity constant, $\eta_0$, can be based on experiments or set such that the viscosity at a reference present day mantle temperature matches constraints on present day mantle viscosity. With initial conditions, the model is then closed and equation (\ref{Tdot}) can be integrated forward over time. \\

A $\beta$ value of 0.33 is valid if internal mantle viscosity is the dominant resistance to the motion of tectonic plates \citep[e.g.][]{Davies1980,Schubert1980}. If plate strength offers significant resistance then the scaling constant has been argued to be 0.15 or less \citep{Christensen1984,Christensen1984a,Christensen1985,Conrad1999}. More recently, \citep{Korenaga2003} has argued that $\beta$ is negative as a result of dehydration during plate formation (plates become stronger in the Earth's past and, as a result, plate velocity and associated mantle cooling decreases even though $Ra$ increases). To account for these different assumptions we will examine cases with $\beta$ values of 0.33, 0.15, 0.0 and -0.15. It is worth noting that variations in $\beta$ represent different assumptions regarding physical processes that are critical to planetary cooling. This distinguishes $\beta$ variations from uncertainties in material parameters (e.g. viscosity function parameters) and initial conditions. Stated another way, variable $\beta$ is associated with competing hypotheses regarding planetary cooling. \\

There is a further break between models that have assumed negative or very low $\beta$ values and classic thermal history models (CTM). As noted, CTM have traditionally integrated the energy balance equation forward in time. Workers who argued for low or negative $\beta$ have taken a different approach and used the present day as the starting point and integrated backwards in time \citep{Christensen1985,Korenaga2003}. A key thought behind that approach is the idea that present day observations, i.e. data constraints, have the lowest data uncertainty (i.e., lower than data constraints for past conditions). The constant $a'$ in equation (\ref{Q}) is set such that present day heat flux ($Q_0$) is achieved for assumed present day mantle temperature and viscosity values ($T_0$ and $\eta (T_0)$, respectively). The present day ratio of heat produced within the mantle to that which is transferred through the surface, termed the Urey ratio ($Ur$), can then used to set present day radiogenic heat production according to $H_0=Ur*Q_0$. The thermal history model then combines equations (\ref{Tdot}), (\ref{H}) and (\ref{Q}) resulting in
\begin{equation} \label{Ebal}
C\dot{T}=UrQ_0\sum_{n=0}^{4} h_nexp\left(\lambda_nt\right)-Q_0\left(\frac{T_m}{T_0}\right)^{1+\beta}\left(\frac{\eta (T_0)}{\eta (T_m)}\right)^{\beta}.
\end{equation}
Equation (\ref{Ebal}) is then integrated backwards in time to produce the Earth's thermal history constrained by present day values of mantle temperature, heat flux, and Urey ratio (we will refer to this as an Earth-Scaled Model (ESM) formulation).

To allow for an apples to apples comparison of uncertainty accumulation for a given $\beta$ model, we use the ESM formulation solved forwards in time. To accommodate for this, and to test low and negative $\beta$ models in a manner that holds to the methods of the workers who argued for these models, we first solved the energy balance equation backwards by prescribing $T_0$, $Q_0$ and present day $Ur$. Integrating backwards to 4500 Ma, using a Fourth-Order Runge-Kutta scheme, provided the mantle temperature ($T_i$) shortly after accretion and differentiation. The value of $T_i$ was then applied as the initial condition for forwards integration for a period of 4500 Myr. Following this procedure, forward integration recovered the initial conditions (i.e., present day values) of the ESM backward integration approach. This then allows for an apples to apples comparison to evaluate how different models respond to perturbations applied forward in time. \\

\subsection{Perturbations}
\indent A useful metric, for dynamic systems models, is their time response - the time it takes a singular perturbation to decay \citep{Close2002,Seely1964}. This metric is referred to as the system's reactance \citep{TexasInstrumentsIncorporated.LearningCenter.1978}. The reactance for variable thermal history models can be gauged by applying a single perturbation to mantle temperature and tracking the time it takes the thermal history path to damp the change in temperature. In addition to being a useful metric on its own, this will also provide insight into potential model uncertainty associated with unmodeled effects that are not singular in time. If the unmodeled effects are associated with variations that would occur on a timescale longer than the system's reactance time, then the system has the potential to damp the variations. If the variations occur on a timescale that is shorter than the systems reactance time, then the variations could be amplified over time and a model has the potential to lose structural stability.

We will perform single-perturbation analysis on full thermal history models and on stripped down versions that exclude the effect of decaying heat sources. The full model analysis will give metrics for full system reactance while the later will give metrics for convective reactance for different models. To isolate the influence of the convective reactance, Equation \ref{Ebal} was modified by dropping the heat source term from the right hand side. Using the modified equation, a reference solution was generated by defining an initial temperature and then allowing the mantle to evolve towards colder temperatures. The initial temperature was then perturbed by 15 $^oC$ to generate a perturbed solution. The difference between these two cooling paths was used to evaluate the convective reactance. If the two paths were converging, the perturbation was decaying, and for a decaying perturbation, an e-folding time was defined as the amount of time it took for the perturbation to decrease by $1/e$. A divergence between the reference and perturbed solutions indicated perturbation growth. For perturbation growth, the e-folding time was defined as the amount of time it took the perturbation to grow by a factor of $e$. The rate at which perturbations grow or decay, for full and stripped down models, provides insights into how models will respond to unmodeled effects and on the potential that specific models may be prone to structural instability. \\

This singular in time perturbation concept can be extended by adding a low amplitude noise term to the governing equation(s) of a model. This is often referred to as a "perturbed physics" analysis. After assessing the reactance times of different thermal history models, we perform uncertainty and stability tests by adding randomized, low amplitude perturbations to each model. Physically, this additional term will mimic the chaotic nature of mantle convection - a factor that is not included in parameterized thermal history models. As well as assessing structural stability, these tests will give metrics for comparing the intrinsic uncertainty between models, i.e., models whose outputs change by a small percentage, due to low amplitude noise, will have a lower intrinsic uncertainty relative to models whose outputs change by a larger percentage.  \\

For the randomized perturbations, the total distribution, for each model integration, was assumed to be normally distributed according to
\begin{equation}
f\left(x|\mu, \sigma\right)=\frac{1}{\sqrt{2\pi\sigma^2}}e^{-\frac{\left(x-\mu\right)^2}{2\sigma^2}}
\end{equation}
where $x$ is the percentage by which the current mantle temperature is perturbed and $\mu$ and $\sigma$ the mean and standard deviation characterizing the distribution of perturbations, respectively. During each integration, a perturbation was randomly drawn from the probability density function (PDF) where $\mu=0$ and $2\sigma=A$. The parameter $A$ is the prescribed amplitude that defines the positive and negative limits between which 95$\%$ of perturbation percentages are located. Therefore, the density of large impulses will be much less than the density of smaller impulses. The impulses are applied at a fixed time interval (the interval itself can be varied).

For the case of randomized perturbations, different perturbation time series are possible. For those cases, multiple integrations were performed, for each model, to simulate the manner in which differing random draws affected the uncertainty and stability of different models. For each specific thermal history model, i.e. models with different $\beta$ values, the mean and standard deviation of $T_m$ were calculated at each time step to test the reproducibility of the average mantle temperature that the thermal history models track and to assign associated model uncertainty windows. \\

\subsection{Coupled Thermal History and Deep Water Cycling Models}
\indent  As previously noted, an advantage of parameterized thermal history models is that added complexity can be incorporated into them relatively efficiently to build models that couple a range of planetary processes (e.g., models that couple the geologic history of a planet to its climate history). Adding complexity has the potential to transform a structurally stable model into an unstable one (or an unstable model into a stable one). Whether this will or will not be the case is not always easy to assess \emph{a priori} as added complexity in planetary models can introduce multiple new feedback loops. Some of the loops may be negative (which would favor stability) but some may be positive (which could allow for instability).  As an example, we will assess the intrinsic uncertainty of a model that couples deep water cycling to thermal history. \\

The full description of the deep water cycling model can be found in Sandu et. al. [2011]. A conceptual sketch will be useful at this stage. Warm mantle rises beneath mid-ocean ridges (MOR) and is subjected to decompressional melting if it passes the mantle solidus, which is depressed by the presence of volatiles (e.g. water). In the volume of melt produced, a fraction is hydrated resulting in a net dehydration of the mantle and therefore an increase in mantle viscosity. This can lower the vigor of mantle convection which can lead to a hotter mantle and, hence, increased melting and dehydration. This allows for the potential of a positive feedback. As the melt carrying the water migrates to the surface, some of the water is locked into the lithosphere whereas the remaining water escapes to the hydrosphere. As the lithosphere is advected away from the MOR, it cools, thickens and gains water via metamorphic processes. At subduction zones, the downwelling slab is heated and releases some of its bound water back to the mantle. The re-injection of water into the mantle lowers its viscosity. This enhances cooling and cooler conditions favor enhanced mantle rehydration at subduction zones. This allows for a feedback that works against the previously noted feedback.  The degree to which the feedbacks are or are not balanced, at any time in model evolution, will determine a models uncertainty metrics. This can be assessed following the procedures outlined in the previous section. An advantage of running an analysis on the base level thermal history model and the water-cycling version is that the effects of different model components on the final intrinsic uncertainty can be assessed. \\

\begin{table}
\caption{Convection Parameters}
\centering
\begin{tabular}{ c c c c c }
\hline\hline
 & Model & Parameter & Value & Units    \\
\hline
Fixed model parameters & & & & \\
 & & $\eta_0$ & $2.21*10^9$ & Pa s          \\
 & & $A$      & 300  & kJ mol$^{-1}$        \\
 & & $R$      & 8.314  & J K$^{-1}$ mol${-1}$ \\
 & & $Q_0$    & 36     & TW \\
 & & $T_0$    & 1623   & K  \\
Model dependent parameters \\
 % & \emph{Korenaga, 2003} \\
 & $\beta=-0.15$ & & & \\
 & & Ur    &  $0.15$  & - \\
 & & $T_i$ &  $1956$  & K \\
 %& \emph{Christensen, 1985} \\
 & $\beta=0.00$ & & & \\
 & & Ur    &    $0.35$  & - \\
 & & $T_i$ &  $1939$  & K \\
 %& \emph{} \\
 & $\beta=0.15$ & & & \\
 & & Ur    &  $0.55$  & - \\
 & & $T_i$ &  $2182$  & K \\
 %& \emph{} \\
 & $\beta=-0.15$ & & & \\
 & & Ur    &  $0.75$  & - \\
 & & $T_i$ &  $2016$  & K \\
\hline
\end{tabular}
\label{table:convectiion_parameters}
\end{table}

\begin{figure}[!h]
    \centering
    \includegraphics[width=0.75\textwidth]{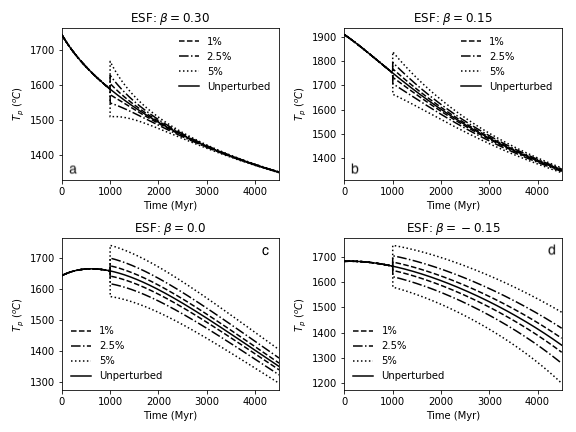}
    \caption{Thermal history response to single in time perturbation of variable frequency. The unperturbed reference model for each model is the solid black line. The different choice of beta for each model represents different model assumptions. As $\beta$ is decreased from its classic value to zero and then into the negative domain, it takes longer for convective processes to eliminate the perturbation, if it does so at all.}
    \label{fig:Fig1}
\end{figure}

\subsection{Exoplanet Scaled Models (XSM)}
\indent The discovery of large terrestrial exoplanets (LTePs) has lead to Earth-based thermal history models being scaled for larger mass and volume planets \citep{Valencia2006,Schaefer2015}. Scaling a model has the potential to alter its uncertainty metrics. As an example, we will scale two models of the previous sub-sections (a water-cycling and a negative $\beta$ model) and assess the intrinsic uncertainty of the scaled models. We will follow the scaling approach of Valencia et al. [2006]. The scaling holds core mass fraction constant at a value of 0.3259. Planet and core radius scale as $R_i \sim R_{i,\oplus} \left(\frac{M_p}{M_\oplus}\right)^{a_i} $ where $\oplus$ indicates the value of Earth. If the index  i represents planet radius, then  $a_p $ = 27. If the index  i represents core radius, then $a_c $ = 0.247. From these constraints, mantle volume and therefore average mantle density, $\langle \rho_m \rangle$ can be calculated. The acceleration due to gravity for each planet is defined by the relationship $\frac{GM_p}{R_p ^2•} $. For a scaled water cycling model, ridge length ($L$) is assumed to scale as 1.5 times the planetary radius. The concentration of mantle water is held constant. This means that total water in the mantle will not be constant. We hold all other parameters fixed and compute thermal evolutions for 1 M$_\oplus$ and 2 M$_\oplus$. \\

\section{ANALYSIS}
\indent In this section, we first assess the structural stability and intrinsic uncertainty of several thermal history models using singular in time perturbations and then low amplitude random perturbations that mimic the chaotic nature of mantle convection. Next, we assess the uncertainty of a model that adds a layer of complexity to a particular base level thermal model to evaluate the influence of competing feedback structures on the evolution of model uncertainty. At the end of this section is an uncertainty analysis of two models, both argued to be applicable to Earth, scaled for LTePs. \\

\subsection{Impulse Response and Reactance Time Analysis}
\indent Figure \ref{fig:Fig1} shows the response of several thermal history models to a singular in time perturbation. All of the models have been scaled to match the same current day mantle potential temperature. Model parameters can be found in Table \ref{table:convectiion_parameters} The perturbation decays most rapidly in the classic thermal history model ($\beta = 0.3$) signifying a short reactance time. This relates to something that has long been noted about such models: results for present day values are insensitive to different initial condition assumptions \citep{Davies1980,Schubert1980}. Classic thermal history models assume that the resistance to the motion of tectonic plates, and associated mantle cooling, comes from mantle viscosity. This, together with the exponential dependence of mantle viscosity on temperature \citep[e.g.][]{Kohlstedt1995}, allows for a strong negative feedback in the model system \citep{Tozer1972}: if mantle temperature increases/decreases, this decreases/increases mantle viscosity which leads to faster/slower plate velocities, an associated increase/decrease in mantle cooling, and a decrease/increase in mantle temperature that works counter to the initial temperature perturbation. This negative feedback leads to a relatively rapid decay of model perturbations (Figure \ref{fig:Fig1}a). \\

\begin{figure}[!h]
    \centering
    \includegraphics[width=0.75\textwidth]{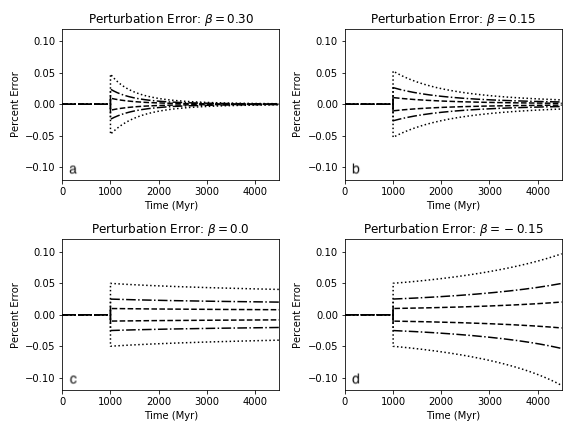}
    \caption{The residual between the perturbed and reference model shows how quickly each model reacts out the single perturbation. (a-c) As $\beta$ is reduced, the time it takes for the perturbation to decay by 1/e, a measure of the reactance time ($\tau$), increased. (d) When $\beta=-0.15$, the perturbation grows rather than decays.}
    \label{fig:Fig2}
\end{figure}

As $\beta$ was decreased, reactance time increased. Models with decreasing $\beta$ assume that plate motion is resisted by a combination of plate strength and mantle viscosity \citep{Conrad1999}. This weakens the negative feedback discussed above and, as a result, model reactance time increases (Figure \ref{fig:Fig1}b). The assumption for $\beta=0$ models is that resistance to plate motion is dominated by plate strength which has no dependence on mantle temperature \citep{Christensen1984,Christensen1984a,Christensen1985}. This removes a negative (buffering) feedback from the model system and the effects of perturbations become long lived (Figure \ref{fig:Fig1}c). The slow convergence of perturbed model paths in Figure \ref{fig:Fig1}c is due to the decay of mantle heat sources - over a long evolution time, heat sources will be tapped and all the paths must approach a final non-convecting state (the model analog for a geologically dead planet). Models with negative $\beta$ assume that resistance to plate motion is dominated by plate strength and further assume that plate strength increases with mantle temperature \citep{Korenaga2003,Korenaga2008}. This leads to a positive feedback within the model system \citep{Moore2015}.  As a result, model evolutions become highly sensitive to assumed initial conditions \citep{Korenaga2016}. In addition, evolution paths become highly sensitive to perturbations. The existence of a positive feedback in the model system means that small perturbations, resulting from any unmodeled effect(s), will be amplified over model evolution time.   \\

The differences in model reactance times can be highlighted by tracking the percent error between the perturbed and reference model paths (Figure \ref{fig:Fig2}). The percent error decayed quickly for models with shorter reactance times. Increasing reactance times, caused by shifting $\beta$ towards zero, led to a slower decay. The positive feedback, associated with a negative $\beta$ model, caused percent error growth. For any given $\beta$, the decay or growth rate in percent error was a function of perturbation sign. Warmer perturbations decayed faster than colder ones for all models with $\beta\geq 0$. \\

\begin{figure}
    \centering
    \includegraphics[width=0.5\textwidth]{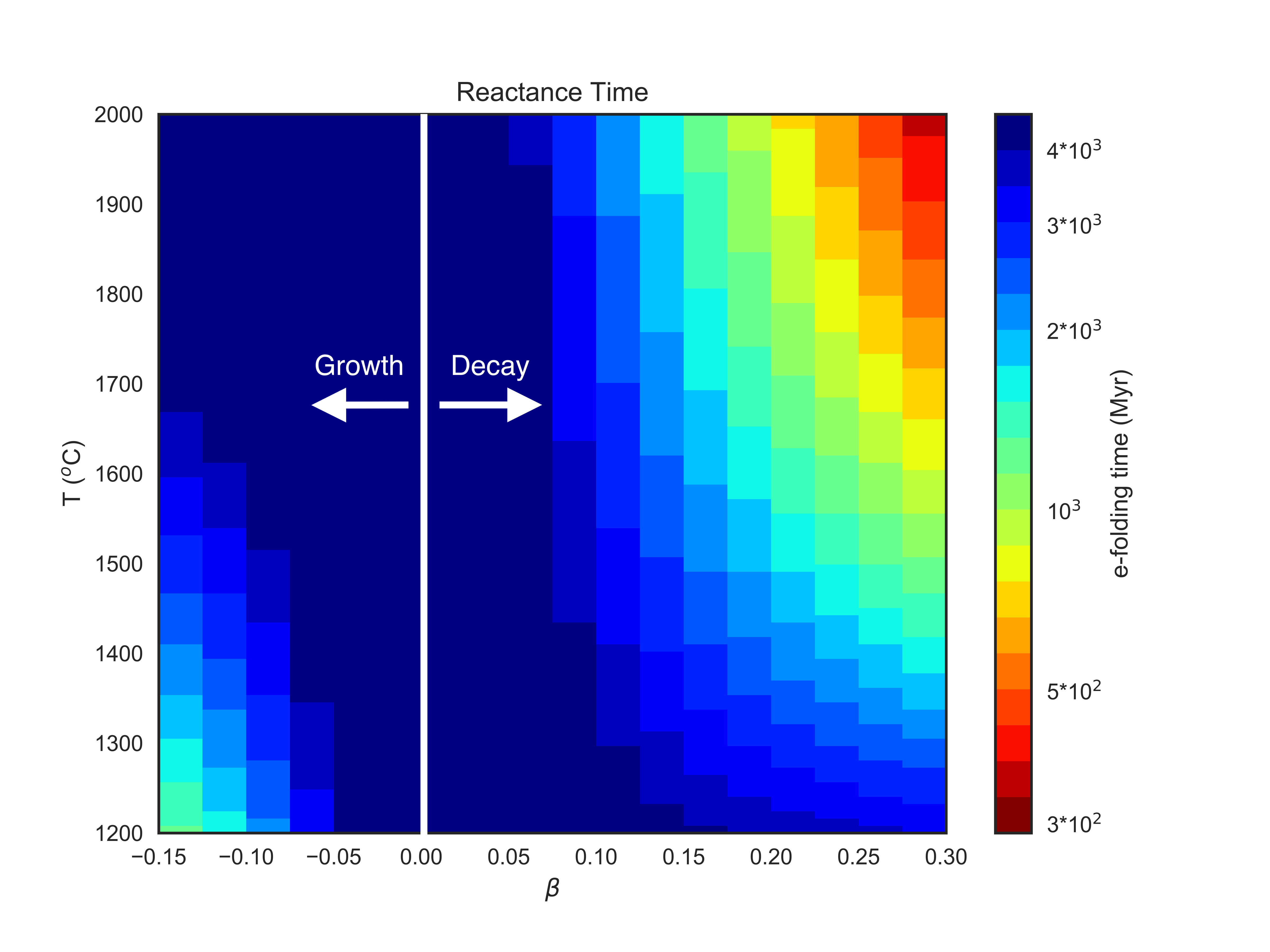}
    \caption{The reactance time of the convective system ($\tau_c$) is a function of temperature (\emph{T}) and the choice of $\beta$ and perturbation amplitude (fixed at 15 $^oC$ here). For $\beta>=0$, the perturbation was damped by the model's negative feedback. A stronger negative feedback (more positive $\beta$) as well as increased T lowered $\tau_c$. For $\beta<0$, the model's positive feedback, which strengthened as $\beta$ decreased, amplified the perturbation. The close $\beta$ was to zero, either positive or negative, the more likely $\tau_c$ was to be greater than the age of the Earth.}
    \label{fig:Fig3}
\end{figure}

The asymmetry between warm and cold perturbations indicates that reactance time is a function of $\beta$ and of mantle temperature. To assess the influence of these two effects, we isolated the convective reactance time ($\tau_c$), defined as an e-folding timescale for perturbation growth or decay (this metric removes the effects of decaying heat sources on model reactance times). The value of $\tau_c$ was calculated for different initial temperatures and choices of $\beta$ (Figure \ref{fig:Fig3}). For $\beta>0$, the system is governed by a negative feedback. When a positive perturbation is added to the system, convective vigor and heat transport also increase relative to the unperturbed state, leading to a rapid decay in the perturbation. When a negative perturbation is added to the system, convective vigor and  heat transport decrease relative to the unperturbed state. The overall feedback remains negative which still leads to a perturbation decay but at a slower rate than for a positive perturbation. As the negative feedback was weakened, by decreasing $\beta$, $\tau_c$ increased. The lengthening of $\tau_c$ continued until $\beta$ took on negative values. 

When $\beta$ is reduced past zero, the overall model feedback changes from a negative to a positive. For negative $\beta$, $\tau_c$ is interpreted as the amount of time it takes a perturbation to grow by a value of $e$. As $\beta$ became more negative, perturbations grew more rapidly (Figure \ref{fig:Fig3}). For a positive feedback, a perturbation that increases mantle temperature decreases the convective efficiency and reduces convective heat transfer. Decreasing mantle temperature leads to the opposite effect which resulted in a greater divergence between the perturbed and the reference cooling path, i.e., an enhanced perturbation growth. The fastest perturbation growth occurred at the combination of coldest temperatures and most negative $\beta$ values (Figure \ref{fig:Fig3}). At this combination, the positive feedback is strongest. In general, a greater positive model feedback will lead to an enhanced amplification of unmodeled affects. This is qualitatively intuitive. The added value of the analysis is that it provides quantitative measures of model uncertainty for different models along different points in their evolution trajectories. \\

\subsection{Stochastic Fluctuations and Perturbed Physics Analysis}
\begin{figure}
    \centering
    \includegraphics[width=0.75\textwidth]{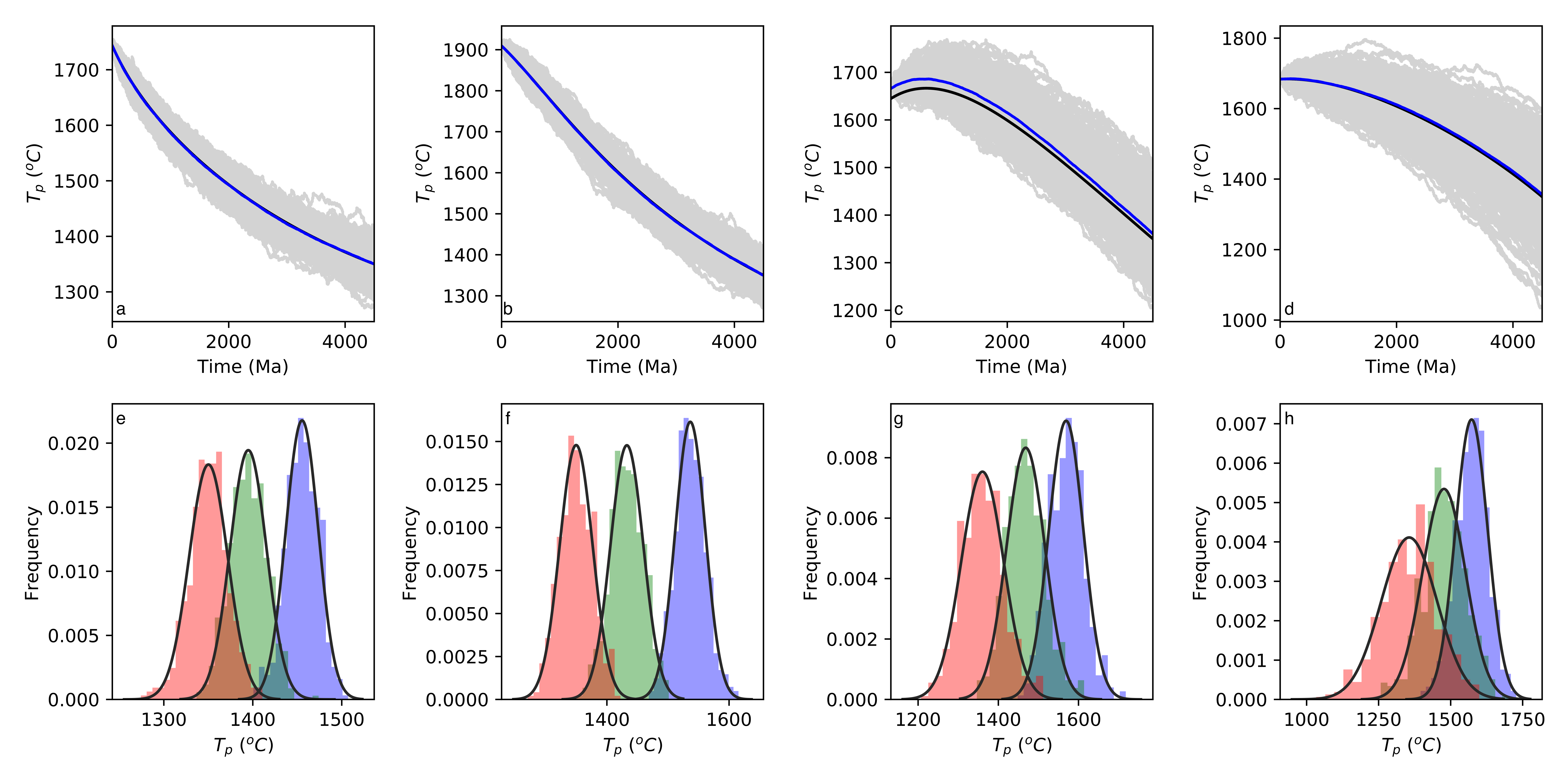}
    \caption{Randomly perturbing the model at a fixed time interval during the thermal evolution results in a cloud of evolution paths, the gray lines in a-d, about the mean (black lines). The blue lines in a-d are the statistical mean of the clouds and match fairly well with the reference model. As $\beta$ was decreased, the uncertainty cloud widened. Distinct time slices through this cloud -- red, green and blue are 0, 1000, and 2000 Ma -- result in normally distributed temperatures with increased variance as the model evolved towards present day as $\beta$ decreased.}
    \label{fig:Fig4}
\end{figure}

The response of thermal history models to random perturbations, over model evolution time, is shown in Figure \ref{fig:Fig4}. Each model was subjected to 500 random perturbation sequences. During each sequence, the model was perturbed at a 10 Myr interval. Each perturbation was drawn from a distribution defined by a mean of zero and a standard deviation of 7.5 $^oC$. The maximum perturbation was then approximately 1\% of the total temperature, considerably less than the natural variation due to the chaotic nature of vigorous convection \citep[e.g. figure 4 of][]{Weller2016}. Plotting all of the perturbed paths generates an uncertainty shadow (top row of Figure \ref{fig:Fig4}). The uncertainty shadow is plotted along with the mean trend from the 500 perturbed models (blue lines) and the trend from the unperturbed models (black lines). A comparison of the mean trends and the unperturbed model trends shows that all the models maintain structural stability. All the models are not, however, associated with the same uncertainty structure. For the classic thermal history model, model uncertainty tended to saturate tightly around the reference solution. As $\beta$ was decreased, model uncertainty increased.  \\

For any specific model evolution time, a probability distribution of mantle temperatures can be compiled. Distributions for the present day, 1000 Ma and 2000 Ma are shown in the bottom row of Figure \ref{fig:Fig4}. Each distribution is normally distributed about the unperturbed path with a time-dependent standard deviation. The smaller the standard deviation, the lower the model uncertainty. The classic model had the tightest distribution about its mean, with slight accumulation of uncertainty over model time. As $\beta$ decreased, the probability distributions became less certain and uncertainty accumulated for longer evolution times. \\

\begin{figure}
    \centering
    \includegraphics[width=0.75\textwidth]{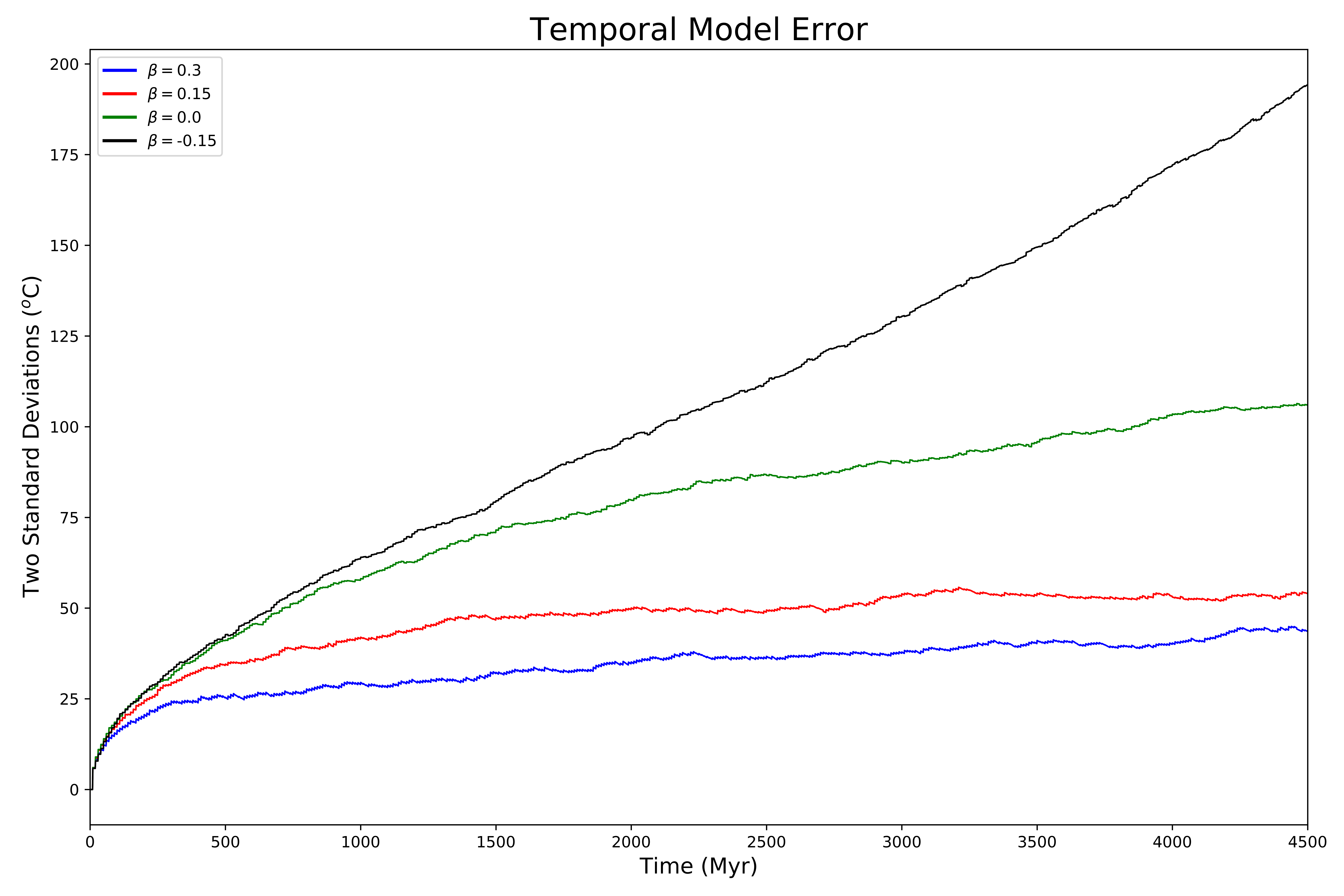}
    \caption{The bound of the uncertainty cloud (two standard deviations from the mean) shows that models with lower reactance times unsurprisingly accumulate less uncertainty. Over the age of the Earth, a model with a positive feedback tended to have uncertainty saturate whereas a negative feedback model accumulated more uncertainty.}
    \label{fig:Fig5}
\end{figure}

Rather than limit the analysis to three specific model times, a continuous uncertainty window can be tracked (Figure \ref{fig:Fig5}). The choice of $\beta$ sets the sign and strength of the principal model feedback. For positive $\beta$, a negative feedback buffers the system and limits the growth of model uncertainty. The uncertainty saturation limit is proportional to $\tau_c$. Recall, for $\beta>0$, $\tau_c$ is a measure of how fast a perturbation decays. A small $\tau_c$ indicates that the system reacts out perturbations quickly, resulting in a smaller accumulation of uncertainty. As $\beta$ approached zero, but remained positive, $\tau_c$ increased, which enhanced uncertainty accumulation, though it remained bounded. In short, positive $\beta$ models tend to accumulate uncertainty to a limit that is proportional to the $\tau_c$ value of the model. \\

Reducing $\beta$ below zero shifted the dominant system feedback from negative to positive. Unlike the $+\beta$ models, there is no uncertainty saturation limit for $-\beta$ models. Instead, as the models evolved to lower temperatures, there was an increase of model uncertainty and an acceleration in its accumulation. As the mantle secularly cooled, any positive perturbation took the perturbed model further from the reference model, reducing the rate at which it cooled, creating a greater divergence between the two models. Alternatively, any negative perturbation created more rapid cooling in the perturbed model, causing it, too, to diverge from the reference model. In either scenario, the cooler of the two model paths was always moving towards colder mantle temperatures more rapidly than the warmer model. In short, the rate at uncertainty accumulated for negative $\beta$ models scaled inversely with the $\tau_c$ value of the model. \\

\subsection{Thermal-Volatile Evolution Model}
\begin{figure}
    \centering
    \includegraphics[width=0.75\textwidth]{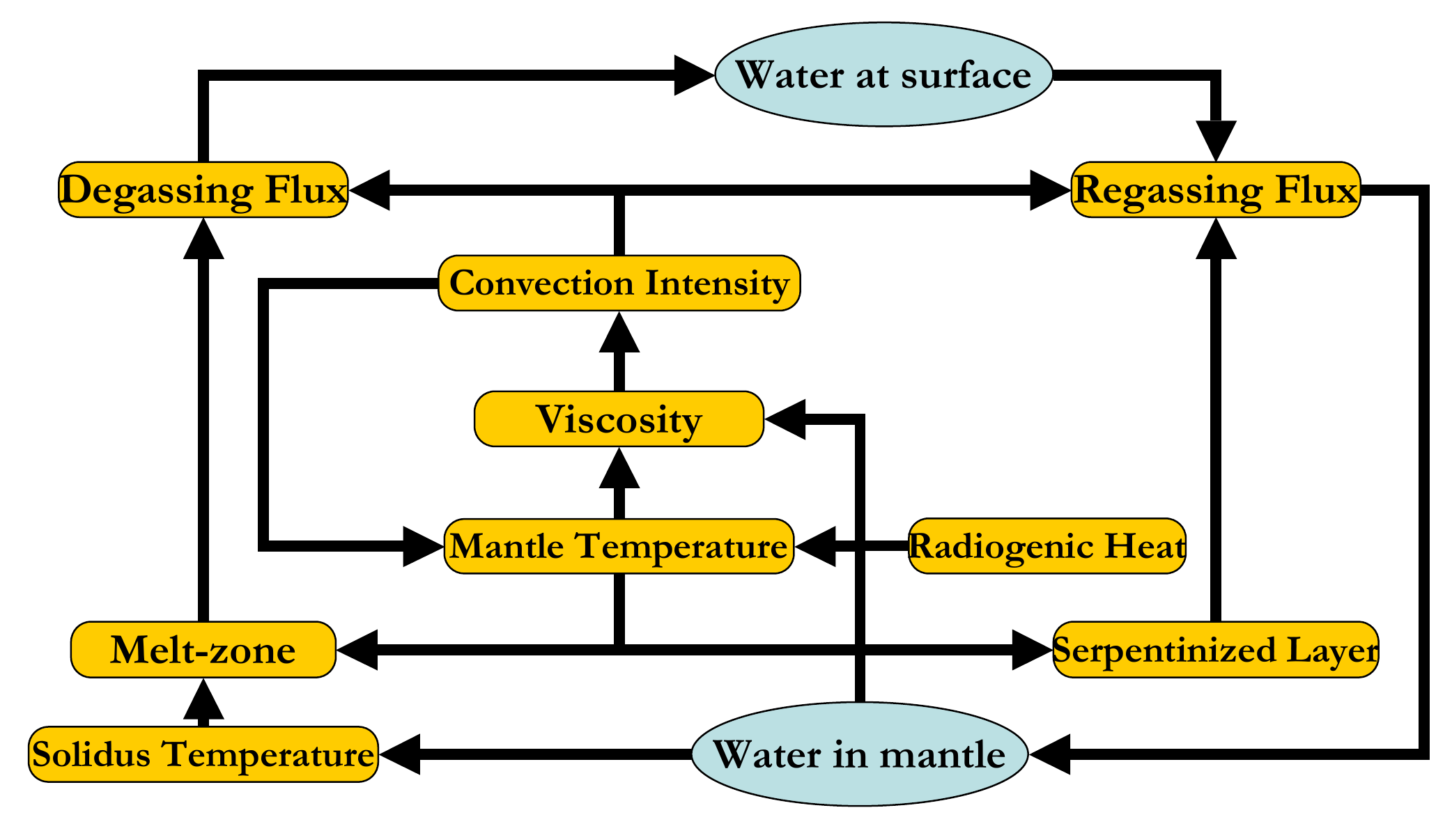}
    \caption{By complicating the thermal history model, accounting for how dehydration of the mantle affects its viscosity, is governed by complicated feedback structures. Within these structures, it may not be obvious if the model's dominant feedback is positive or negative, such as in the competition between temperature (negative feedback) and water effects (positive or negative feedback) on the mantle visocisty.}
    \label{fig:Fig6}
\end{figure}

\indent Introducing new layers of complexity to a model has the potential to change the stability and uncertainty properties of the model. As an example we will consider adding deep water cycling to a thermal history model [see \citep{Sandu2011} for details]. Figure \ref{fig:Fig6} illustrates the feedbacks introduced by coupling deep-water cycling to thermal history. Mantle viscosity depends inversely on mantle temperature and water concentration. Mantle temperature also influences deep water cycling and, by association, the concentration of water in the mantle at any given time. At mid-ocean ridges, a warmer mantle depresses the solidus, increasing the melt zone thickness and the volume of water-rich melt. This melt ascends to shallower depths and water is released to the surface reservoir, dehydrating the mantle. Lowering mantle water content increases mantle viscosity. This can reduce convective vigor which, in turn, can drive mantle warming. This allows for a positive feedback in the system, even for positive $\beta$ models (the more the mantle is dehydrated, the warmer it will be, and the more dehydrated it will become). Mantle temperature also influences the amount of water brought back into the mantle at subduction zones. Water is stored in the lithosphere as serpentinites to the depth of the 700 $^o$C isotherm \citep{Ulmer1995}. A thicker lithosphere will have a greater volume of water trapped in the slab. The thickness of the downwelling slab is set by the mantle temperature. The mantle de- and rehydrating processes, outlined above, coupled together with mantle temperature to determine mantle viscosity which, in turn, feeds into convective vigor. As a result, model reactance times, structural stability, and intrinsic uncertainty can all be altered due to the introduction of deep water cycling. \\

\begin{figure}
    \centering
    \includegraphics[width=0.5\textwidth]{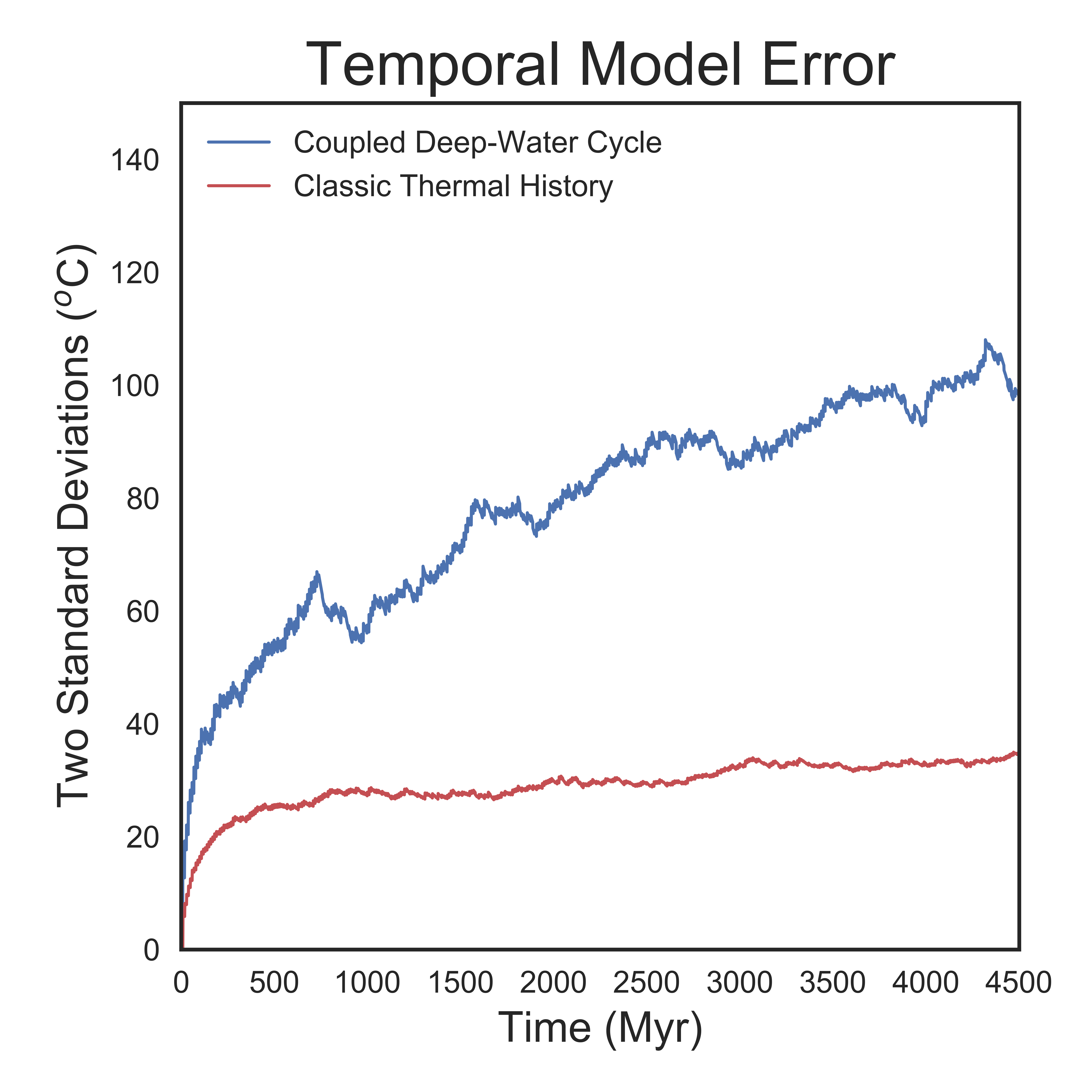}
    \caption{The deep water cycle model (blue line) weakened the negative feedback in the model, lengthening $\tau_c$, which allowed for more uncertainty accumulation as compared to the same thermal history model without water cycling (red line).}
    \label{fig:Fig7}
\end{figure}

Uncertainty was evaluated for the coupled model using the procedures of the previous subsection. The uncertainty shadow from the coupled model is plotted in Figure \ref{fig:Fig7} along with that from the classic $\beta$ reference model. Uncertainty was almost tripled relative to the reference case at the final model evolution time. The addition of a positive feedback associated with the deep water cycle decreased the overall negative feedback of the model. Although the uncertainty is not fully bounded over the model run time shown in Figure \ref{fig:Fig7}, it is approaching a limit asymptotically (this will be made clear in the next subsection). Enhanced uncertainty in the coupled model, relative to a base level thermal model, is due to the continuous competition between thermal and water cycle effects on mantle viscosity.The competition can alter the overall system feedback over model evolution time, which effects the accumulation of uncertainty. This points to the value of performing simplified, analytic feedback analysis to map possible model trends consistent with internal feedbacks \citep[e.g.][]{crowley2011,Astrom2008}. Feedback analysis can provide qualitative insights into uncertainty potential (e.g., does the model allow for unbounded uncertainty) without having to run the full model. For models of the type explored here, the computational time to run models is not a restriction but for more complex models it can be which adds to the value of feedback analysis as a first step in evaluating uncertainty potential.  \\

\begin{figure}
    \centering
    \includegraphics[width=0.75\textwidth]{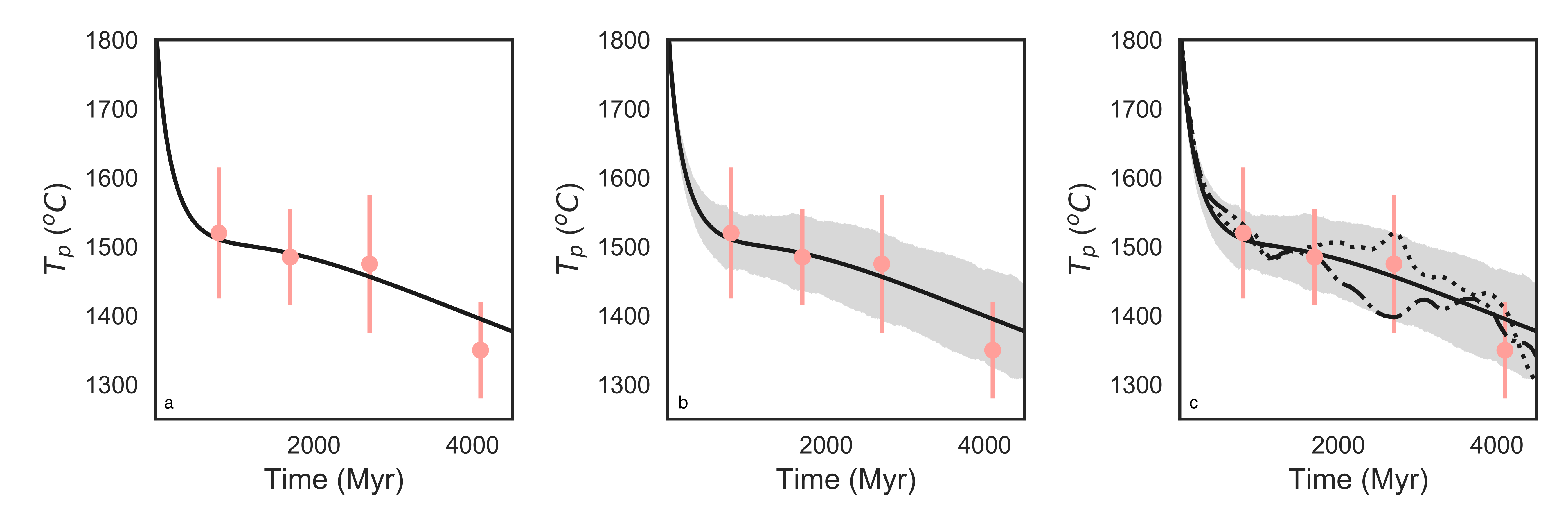}
    \caption{Comparison of a model is a way of testing how successful it is. Classically, this means plotting the thermal evolution to see how well it fits the data (a). With the knowledge that uncertainty in the solution exists, we can plot the uncertainty cloud defining the range of solutions along with the mean and compare this with data (b). Within the uncertainty cloud, different cooling trends are capable of satisfying the data, limiting our ability to constrain exactly the path the Earth has taken.}
    \label{fig:Fig8}
\end{figure}

An example of how model results, accounting for intrinsic uncertainty, can be compared with observational data is shown in Figure \ref{fig:Fig8} (the data comes from \citet{Condie2016}). In Figure \ref{fig:Fig8}a, the typical method of presenting a model solution with data constraints is shown. Adding an uncertainty shadow (Figure \ref{fig:Fig8}b) allows for a more complete comparison of the model prediction with observational data. For the chosen initial conditions and parameters, the coupled model could satisfy observational constraints at a model uncertainty level comparable to that of the observational data itself. Two of the model solutions that determined the model uncertainty shadow are shown along with the mean solution in Figure \ref{fig:Fig8}c. Either path, within model uncertainty, is a viable model result. Plotting them shows the level of model deviations that can occur relative to the mean model trend (i.e., the trend from numerous models that account for stochastic fluctuations). The individual paths in Figure \ref{fig:Fig8}c evolved similarly in the initial and final stages of model evolution. However, a peaked divergence of $\sim$150 $^oC$ occurred between them at around 2500 Myr of model evolution. Whether this level of model uncertainty is large enough to alter the results from more complex models that, for example, couple climate evolution to thermal evolution will depend on the dynamic properties of the climate models used. Given that many climate models allow for bi-stable behavior, i.e. multiple solutions under equivalent parameter conditions \citep[e.g.][]{Scheffer2009}, this possibility can not be ruled out. \\

\subsubsection{Large Terrestrial Planet Scaled Model}

Thermal history models are being projected in time as many exo-systems are older than our own solar system. They are also being scaled spatially to model terrestrial planets larger and more massive than Earth. Size scaling and time projection can introduce new layers of uncertainty. There is also uncertainty associated with model selection. Even for the Earth, the planet with the best observational data, there is debate regarding which thermal history model best represents the Earth's thermal evolution \citep[e.g.][]{Conrad1999,Grigne2005,Korenaga2003,Grigne2001,Korenaga2008,Silver2008,Sandu2011,Moore2013} Model selection is a form of structural uncertainty: competing models for the Earths thermal history have different mathematical structures to represent what different researchers consider to be the essential factors that have determined the Earths thermal evolution.  \\

\begin{figure}
    \centering
    \includegraphics[width=0.75\textwidth]{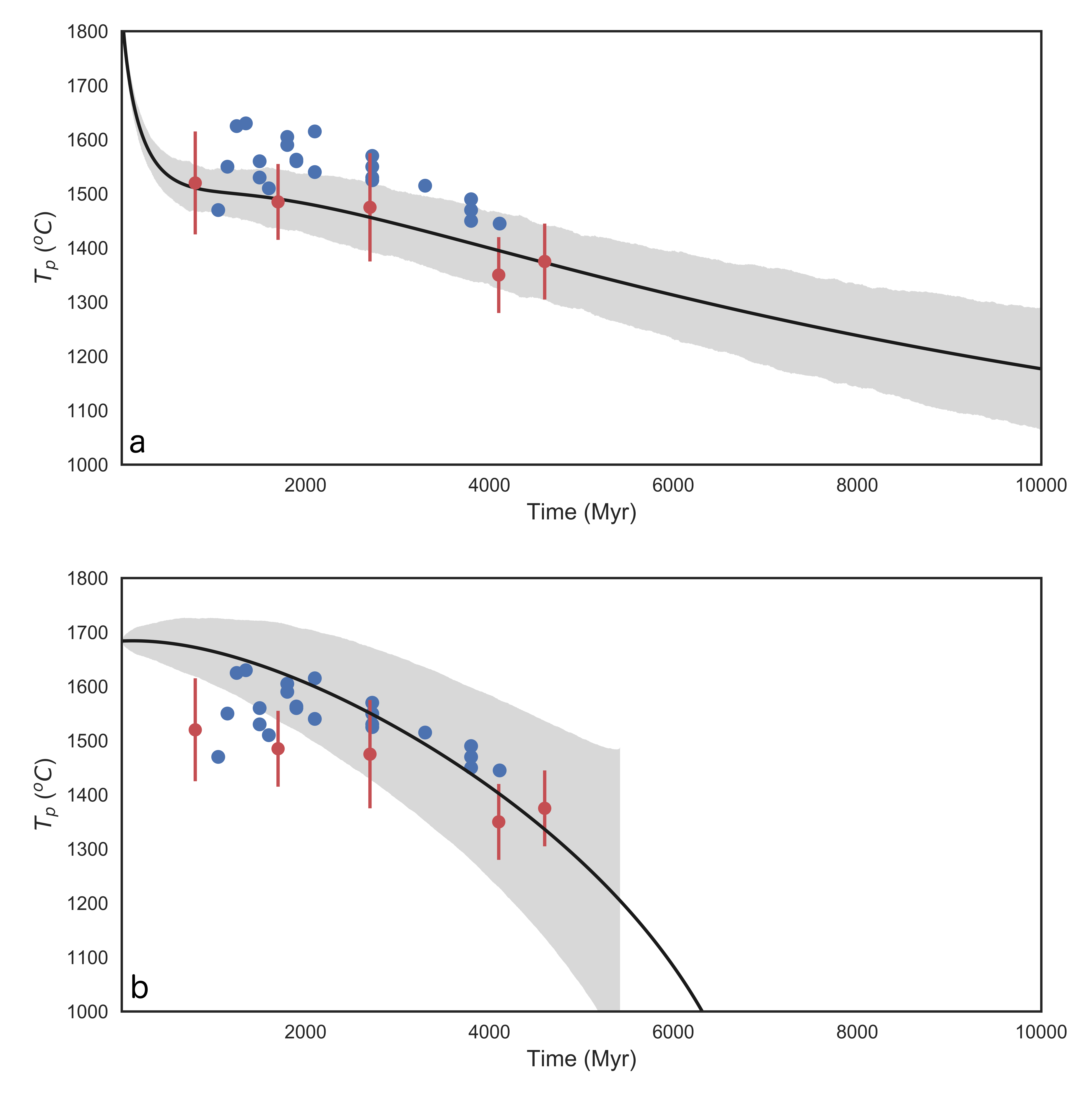}
    \caption{Different models are capable of matching petrologiclal constraints. The dominant feedback in the model, however, has implications in its use for forecasting the thermal history of a terrestrial planet beyond present day, leading to vastly different solutions and whether or not the model becomes unstable.}
    \label{fig:Fig9}
\end{figure}

Figure \ref{fig:Fig9} shows thermal evolution paths from two models that have been applied to the Earth. One model is the deep water cycling model of the previous section \citep{Sandu2011}. The other is an ESM with a $-\beta$ value \citep{Korenaga2003,Korenaga2008}. Petrological data constraints are plotted with each model \citep{Herzberg2010,Condie2016}. The models, as originally presented, did not include uncertainty shadows, which we have determined following the procedures of the previous sections. We have also extended model time beyond the age of the Earth. Although the Earth provides the best observational constraints on a thermal history model, there is still uncertainty in the data. This includes uncertainty in observational constraints on the Earths past thermal state \citep[e.g.][]{Herzberg2010,Condie2016} and uncertainty in observational data that can constrain the Earth's present thermal state (average surface heat flux, present day radiogenic heat production, the average internal temperature of the Earths' mantle \citep[e.g.][]{Jaupart2007,Sarafian2017}. Data uncertainty together model uncertainty (parametric and intrinsic/structural) allows different models to satisfy data constraints within uncertainty windows. Monte Carlo approaches can be used to determine which models can match observational constraints over a larger portion of potential parameter space in an effort to gauge models that are statistically preferred \citep[e.g.][]{McNamara2000,Hoink2013}. Such approaches can rule out some classes of models but multiple competing models still remain viable. In short, models based on different assumptions can fit Earth based data constraints within allowable uncertainty bands. The critical point for modeling exoplanets is that there are competing hypothesis for the the Earth's thermal evolution which brings with it an uncertainty associated with model selection. \\

The model paths plotted in Figure \ref{fig:Fig9} are for particular parameter and initial conditions that can match data trends within uncertainty. The differences in model structure, and associated uncertainty, becomes clear when the models are projected forward in time for 10 billion years (Gyr). The intrinsic uncertainty for the deep water cycling model tends to flatten over time. For the ESM model, uncertainty increases over time and after ~ 5 Gyr of evolution the temperatures on the cold side of the uncertainty shadow become so low that the cold paths rapidly decay beyond that time. This leads to the mean of multiple perturbed models deviating from the unperturbed model trend (for this reason, the uncertainty shadow is cut off at this point - beyond this point, probability distributions of model outputs deviate from guassian and can become fat-tailed on the cold side of model evolution). The ESM model predicts that volcanic-tectonic activity should end after ~5-6 Gyr (mantle temperatures become too cold to allow for continued melt generation). The deep water cycling model predicts a far longer volcanic-tectonic lifetime. If model uncertainty is not taken into account this could lead to some discordant claims about the geologic lifetime of our planet and, by association, about the potential lifetimes of terrestrial planets within an exo-systems.  \\

\begin{table}[!htb]
\caption{Deep Water Cycle Model Parameters}
\centering
\begin{tabular}{ c c c c c }
\hline\hline
Model & Parameter & Description & Value & Units    \\
\hline
Convective Model \\
& $T_s$ & Surface temperature & 300 & K \\
& H(0) & Initial radiogenic heat & 4.51 & $J/(m^3*yr)$ \\
& Rm & Mantle raidus & 6271 & km \\
& Rc & Core radius & 3471 & km \\
& $\rho_m$ & mantle density & 3000 & $kg/m^3$ \\
& $k_m$ & thermal conductivity & 4.2 & $W(m*k)$ \\
& cp & specific heat & 1400 & $J/(kg*K)$ \\
& $\alpha$ & thermal expansivity & $3.00*10^{-5}$ & $K^{-1}$ \\
& $\beta$ & convective exponent & 0.33 & - \\
& $\lambda$ & decay constant & $3.4*10^{-10}$ & $yr^{-1}$ \\
& $Ra_{cr}$ & critial Rayleigh number & 1100 & - \\
Water Cycling \\
& $\eta_0$ & viscosity constant & $1.7*10^{17}$ & Pa*s \\
& $A_{cre}$ & material constant & 90 & $MPa^{-r/s}$ \\
& r & fugacity exponent & 1.2 & - \\
& $Q_a$ & creep activation energy & $4.8*10^5$ & J/mol \\
& $\chi_d$ & degassing efficiency factor & 0.03 & - \\
& $\chi_r$ & regassing efficiency factor & 0.015 & - \\
& OM & mass of 1 Earth ocean & $1.39*10^{21}$ & kg \\
& OM(0) & ocean masses initially in mantle & 2 & - \\
\hline
\end{tabular}
\label{table:deep_water_cycle}
\end{table}

\begin{figure}
    \centering
    \includegraphics[width=0.75\textwidth]{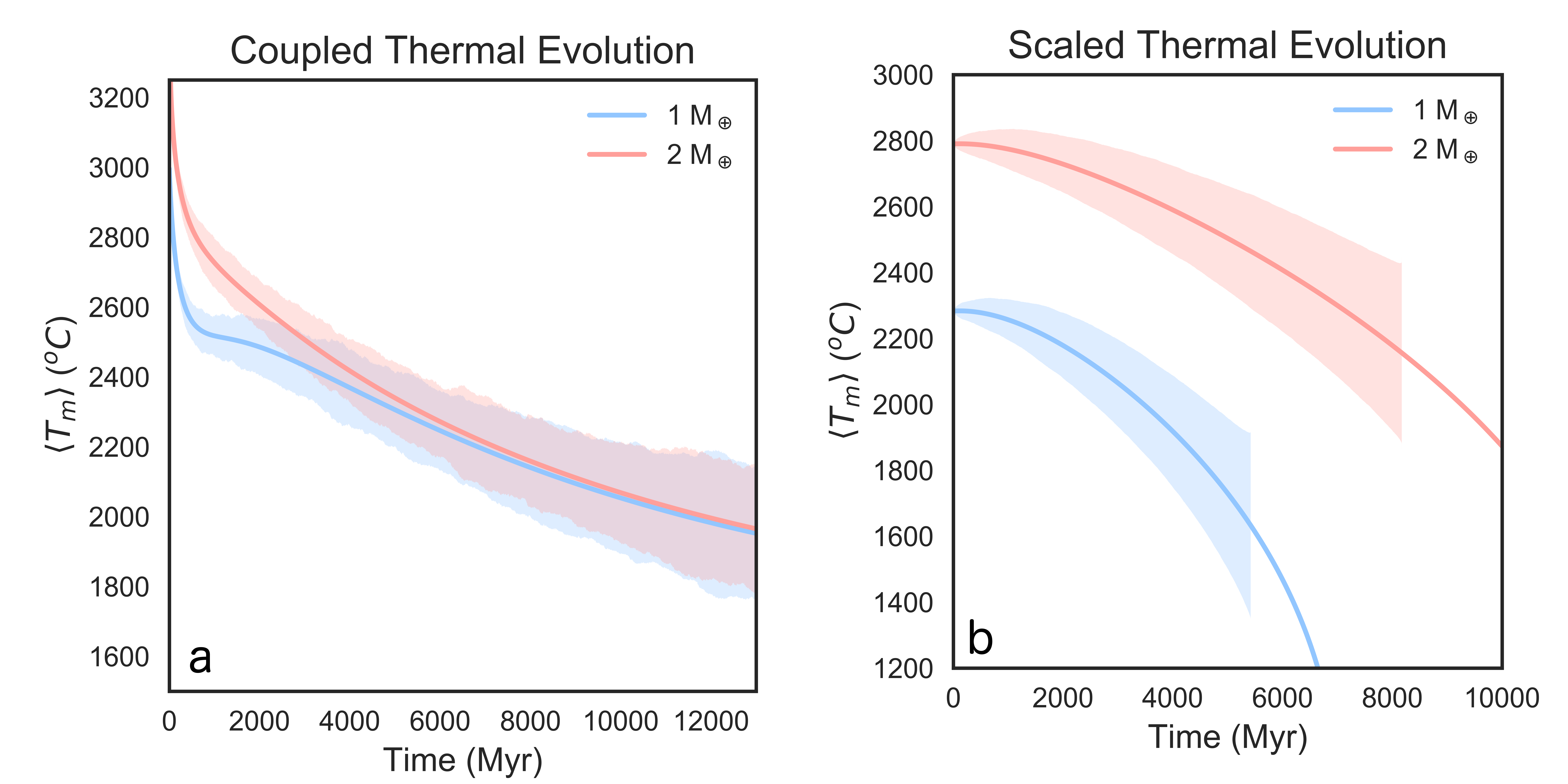}
    \caption{Scaling a model to terrestrial planets of different size may or may not affect the instrinsic uncertainty. In the deep water cycling model (a), since the uncertainty accumulation is dependent on mantle temperature, and the strong positive feedback keeps the model near a qausi-steady state, similar mantle temperatures achieved beyond the age of Earth provide a similar uncertainty estimate. A model with a negative feedback (b) maintains warmer temperatures for longer, preventing the onset of model instability until further into the model evolution}
    \label{fig:Fig10}
\end{figure}

As well as being extended in time, thermal history models can be scaled for larger planets. The simplest method is to increase size and mass while holding all else equal -- an assumption that may or may not be valid and is used herein only for demonstration. Parameters used for the scaled models can be found in Table \ref{table:deep_water_cycle}. For the deep-water cycling models, it is assumed that each planet begins with the same mantle water concentration and the same mantle potential temperature, which is consistent with previous studies that have scaled deep water cycle models \citep[e.g.][]{Schaefer2015}. This implies that after extrapolating along the adiabat to get a temperature depth profile, a larger planet starts with a higher volumetric mantle temperature. In Figure \ref{fig:Fig10}a we plot the reference solutions for $1M\oplus$ and $2M\oplus$ planets together with their respective uncertainty shadows. The model differences that arise in the early stages of evolution result from the competition between thermal and deep-water cycle effects on mantle viscosity. For the choice of parameters, the Earth-sized model experienced a time window of little to no cooling between 1-2 Gyr. During this time, the mantle was dominantly dewatering  which tended to increase mantle viscosity, lower convective vigor, and favor mantle heating (a positive feedback). That heating tendency, due to more sluggish convection, was balanced by the tendency of temperatures to drop due to decaying heat sources. This lead to a flat line cooling trend. Following this period, water cycled back into the mantle from the surface and cooling was enhanced. Doubling the size of the planet increased the strengthen of the negative thermal feedback such that it outweighed the positive feedback due to mantle dewatering (that positive feedback was critical for the flat line cooling phase of the Earth sized model). Both the enhanced convective vigor and reduction of competition between thermal and dewatering effects on mantle viscosity worked to decrease $\tau_c$ for the scaled up model, relative to the Earth model, during the early stages of evolution. This lowered the relative uncertainty of the scaled up model over the first ~2-3 Gyr of evolution. This implies that, for this particular model structure,  the uncertainty of a reference Earth model will not underestimate the uncertainty of a scaled up model. \\

The ESM model with $\beta$ =-0.15 was also scaled for a larger planet (Figure \ref{fig:Fig10}b). Shortly after 5 Gyr of model evolution, the reference model runs away to cooler temperatures. This does not represent an instability in so much as it signals that the evolution has moved outside of the conditions that the model was designed for from the start (models, in general, assume limits of validity and, as such, what they predict should only be physically interpreted within those limits). In this particular case, the model relies on the assumption that melting can occur within the mantle to generate a chemical lithosphere. Once mantle temperature drops so low that melting can not occur the assumed limits of model validity have been crossed. Doubling the mass of the planet increased the resilience against runaway by lessening the positive feedback inherent to this model (cf. Figure \ref{fig:Fig3}). A weaker feedback lead to slower cooling. This effectively increased the $\tau_c$ of the model. As a result, uncertainty grew more slowly than for the Earth model. The scaled up model remained within its limits of validity over the entire 10 Gyr evolution. However, several of perturbed models, that evolved toward the colder side of the uncertainty envelope, did run away at ~8 Gyr of evolution (this is why the uncertainty envelope is shown truncated at ~8 Gyr).  \\

\section{Implications for Planetary Modeling}
There is a saying in the modelling world: all models are wrong but some are useful. Assessing the uncertainty of planetary models is not designed to argue that one model is more or less wrong than another. It is designed to provide insights into how models can be used. Stated another way, uncertainty analysis provides an utility metric, not a validity metric. Uncertainty analysis can not determine if the assumptions of one model are more or less physically valid than those of a competing model. It can not rule out the potential that the Earths' thermal evolution and/or that of terrestrial planets in other solar systems is dominated by processes that lead to an overall positive system feedback and an associated high level of uncertainty. What it can provide are guidelines for how model predictions should be viewed by a modeler and, perhaps more critically, by colleagues who may not be modelers, but who use model results. 

In terms of how thermal history model predictions are viewed, Figure \ref{fig:Fig4} is telling. Model results, for any particular combination of initial conditions and parameter values, are presented as probability distributions. This is not the norm in thermal history modeling to date. Traditionally, any probabilistic notions have entered into thermal history studies due to the fact that initial conditions and input parameter values are not perfectly known. For any combination of specified initial conditions and parameter values, model outputs have been viewed as deterministic. The fact that thermal history models have been argued to be equal if evolved forward or backwards in time shows the strength of the deterministic view \citep{Christensen1985,Korenaga2003}. Once intrinsic uncertainty is accounted for, model outputs/predictions take on a probabilistic element that breaks from strict determinism. The break is not due to any uncertainty in initial conditions and/or parameter values. It is due to the intrinsic/structural uncertainty of the models themselves. The evolution of a particular model, with specified initial conditions and parameter values, is no longer a trajectory but is a cloud of potential solutions. For structurally stable models, the mean of the cloud corresponds to the trajectory of the classic, strictly deterministic view. It is the highest probability path within a model uncertainty window. The shape of the uncertainty window/cloud is determined by the mathematical structure of the model and different models can have similar mean paths but different windows of potential solutions. For structurally stable models, the mean path can be time symmetric but the probability distribution of model outputs will not be the same for a model evolved forward in time and an identical model evolved backwards in time. 

The different view of model outputs, discussed above, effects the way the models can be used. In comparing model predictions to observational data the uncertainty of the data and the uncertainty of a particular model evolution will now both come into play. If a model mean falls within data uncertainty and so too does the uncertianty cloud, then we may have relatively high confidence in that particular model case. If on the other hand, the model uncertainties exceed those of the data, then we must hedge our level of confidence. An extreme case would be one in which model uncertainty far exceeds data uncertainty. Under such a case it would be difficult to use the data itself to rule out particular model paths. In effect, a model could become irrefutable given the data. An irrefutable model is not necessarily based on invalid assumptions but it does loose a level of utility. 

In Earth focused studies, observational data is often used to discriminate between competing thermal history models \citep[e.g.][]{Herzberg2010} and/or to determine which particular parameter combinations are consistent with observations for a specific model \citep[e.g.][]{Sandu2011}. A consideration of intrinsic uncertainty will effect both of these applications. Input parameter combinations that were ruled out based on a strictly deterministic, binary approach (matches or does not match data) may be viable within intrinsic model uncertainty. Once intrinsic uncertainty is accounted for, the question of whether parameter combinations can match data constraints can only be evaluated in a probabilistic sense. The same consideration will hold for evaluating if one model should be rejected in favor of a competing model based on ability to account for observations. If the uncertainty cloud of a model path overlaps observational data (accounting for data uncertainty), then there is no general rule to assess whether the probability overlaps can or can not rule out the viability of the path. The only suggested criteria that holds is that the modeler explicitly state any cutoffs if he or she uses them (e.g., ``we consider parameter combinations that can match data within two sigma confidence, within the model uncertainty window, as potentially viable''). This is not a rule but a practice of clear and transparent communication of model uncertainty. The same criteria holds if a modeler makes the claim of ruling out one model in favor of a competing model. A final issue to be kept in mind, related to the discussion of this paragraph, is that there is no reason why the evolution of a particular planet, e.g. Earth, should match the most probable prediction of a particular model. 

When thermal history models are moved from Earth to exoplanet applications they take on a more predictive, as opposed to postdictive, nature. An application in this realm is using planetary models to map conditions that can maintain liquid water at the surface of a planet over geological time scales, a planetary feature that is considered to be critical for life as we know it [e.g., Meadows and Barnes, 2018]. This requires coupling thermal history models to climate models \citep[e.g.][]{Foley2015,Foley2016,Rushby2018}. This, in turn, means that the models track volatile cycling between a planets interior and its surface envelopes. We have shown how coupling a thermal history model to a volatile cycling model can compound intrinsic model uncertainty in a non-linear way (Figure \ref{fig:Fig7}). Climate models will be associated with their own intrinsic uncertainties \citep[e.g.][]{Mahadevan2010}. How fully coupled thermal history and climate evolution models will propagate and compound intrinsic model uncertainty has not, to date, been evaluated. The approach that has been used acknowledges parameter and initial condition uncertainty and then runs a number of cases to map parameter regions that do or do not maintain surface water. Monte Carlo parameter sensitivity analysis of this sort is of value but if intrinsic/structural uncertainty is large, then the predictions from a coupled model may be highly uncertain even for particular initial condition and parameter values. Several dangerous scenarios are then possible. As an example, a coupled model that does not account for intrinsic uncertainty may determine that a particular parameter combination (e.g. solar distance, planet mass, initial water content) leads to an uninhabitable planet. If intrinsic uncertainty is allowed for, then this conclusion may no longer hold. A second example adds uncertainty in model selection. Coupled climate and thermo-tectonic evolution modeling studies have dominantly used a particular thermal history base level model. Often a model that has been previously applied to Earth is used. The majority of such models have assumed that the tectonic expression of mantle convection for the Earth is plate tectonics - all of the thermal history models tested in this paper make that assumption. This does not remove uncertainty in terms of model selection as shown in Figure \ref{fig:Fig9}. Both models in Figure \ref{fig:Fig9} have been argued to be valid for modeling the evolution of a terrestrial planet operative in a plate tectonic mode. They lead to very different predictions for such planets older than the Earth. Once the possibility that other modes of tectonics could allow for habitable conditions is considered \citep[e.g.][]{Lenardic2016,Foley2018}, then  the need to acknowledge model selection uncertainty, coupled to intrinsic uncertainty, becomes even more pressing. 

The above is not a call for less but for more modeling. Specifically, a layer of modeling that is not initially geared at making predictions that can be compared to data or that can be used to guide future exploration to gather new observations but, instead, is designed to quantify model uncertainties before the models are put into application modes. With this comes the value of a multi-model approach. Different base level thermal history models can be used collectively to determine which conclusions remain robust in light of model selection uncertainty together with the intrinsic uncertainties of particular models. This adds a complication to using planetary models to guide our thinking about conditions that allow for planetary life beyond Earth but the problem is not intractable. The value of parameterized thermal history models, and the coupling of such models to simplified climate models, is that they remain simple enough to run millions of models given current computational power. This can provide uncertainty measures which, together with a multi-model approach, can be used to put confidence limits on planetary model predictions - something that will be appreciated by observationally focused colleagues who have long strived to provide uncertainty measures on observational data. The search for planets that allow for life beyond Earth involves a synergy between modeling and observations \citep[e.g.][]{Kasting2012}. Being open and transparent about uncertainties within each realm is of value in moving the joint venture forward. 

\section{Conclusions}
Thermal history models are associated with uncertainty that is independent of imperfectly known initial conditions and/or input parameter values. This intrinsic uncertainty (i.e., an uncertainty intrinsic to the modeling process) can be assessed. Isolating dominant model feedbacks can give qualitative insight as to whether uncertainty can be amplified or damped. Single perturbation analysis can bring a more quantitative assessment by determining the reactance time of a model (the time for perturbations to grow or decay by some amount, e.g., by a factor of $e$). A perturbed physics approach can provide a further metric by determining an uncertainty shadow for a particular model evolution over time. It can also determine the structural stability of a model and provide model output probability distributions that account for intrinsic uncertainty. Once intrinsic uncertainty is accounted for in thermal history models, model outputs/predictions and comparisons to observational data should be treated in a statistical/probabilistic way.

\bibliographystyle{apalike}
\bibliography{Refs}

\end{document}